# Multiwavelength Mitigation of Stellar Activity in Astrometric Planet Detection

Avi Kaplan-Lipkin[1], Bruce Macintosh[1], Alexander Madurowicz[1], Krishnamurthy Sowmya[2], Alexander Shapiro[2], Natalie Krivova[2], and Sami K. Solanki[2]
[1] Kavli Institute for Particle Astrophysics and Cosmology, Stanford University, California, USA
[2] Max Planck Institute for Solar System Research, Göttingen, Germany

## Abstract

Astrometry has long been a promising technique for exoplanet detection. At the theoretical limits, astrometry would allow for the detection of smaller planets than previously seen by current exoplanet search methods, but stellar activity may make these theoretical limits unreachable. Astrometric jitter of a Sun-like star due to magnetic activity in its photosphere induces apparent variability in the photocenter of order 0.5 m$R_\odot$. This jitter creates a fundamental astrophysical noise floor preventing detection of lower-mass planets in a single spectral band. By injecting planet orbits into simulated solar data at five different passbands, we investigate mitigation of this fundamental astrometric noise using correlations across passbands. For a true solar analog and a planet at 1 au semimajor axis, the 6$\sigma$ detection limit set by stellar activity for an ideal telescope at the best single passband is 0.01 Earth masses. We found that pairs of passbands with highly correlated astrometric jitter due to stellar activity, but with less motion in the redder band, enable higher-precision measurements of the common signal from the planet. Using this method improves detectable planet masses at 1 au by up to a factor of 10, corresponding to at best 0.005 Earth masses for a Sun-like star with a perfect telescope. Given these results, we recommend that future astrometry missions consider proceeding with two or more passbands to reduce noise due to stellar activity.

*Unified Astronomy Thesaurus concepts:* Exoplanet detection methods (489); Astrometry (80); Sunspots (1653); Solar faculae (1494)

## 1. Introduction

To date, over 4000 exoplanets have been discovered by a variety of methods.[3] Most of these have been detected by their transits across the host star. About 900 have been detected by radial velocities instead, using the Doppler shift of the host star due to its orbit around the barycenter of the star-planet system, and about 50 detections have come from direct imaging of planets.

No confirmed planets have been detected by astrometry yet. As the host star and planet orbit their barycenter, astrometric measurements of the host star should oscillate with the orbital period. While astrometry has not yet directly found any exoplanets on its own, such measurements have already helped constrain orbital parameters of planets, such as in the case of 51 Eridani b (Rosa et al. 2015) and β Pictoris b (Nielsen et al. 2020).

One of the earliest suggestions for astrometric planet detection was Project Orion (Black 1980), which suggested use of a ground-based imaging stellar interferometer at visible and infrared wavelengths. Twenty years later, the Space Interferometry Mission was proposed for astrometric detection of Earth-mass planets using Michelson stellar interferometry to achieve high-precision measurements (Goullioud et al. 2008). The mission was canceled in 2010, but it charted out a path toward Earth-size planet detection using interferometry.

With recent improvements in technology and upcoming missions, astrometric planet detection will become possible. The ongoing Gaia astrometry mission (Gaia Collaboration et al. 2016), extrapolated to a 10 yr duration, is expected to find 70,000 new exoplanets (Perryman et al. 2014). These would all be high-mass (approximately Jupiter-mass or larger) planets with long-period orbits. LUVOIR is an upcoming mission expected to find smaller Earth-like planets via astrometry (The LUVOIR Team 2019). For lower-mass planets with smaller orbital periods, astrometric jitter due to magnetic activity in the host star creates a fundamental astrophysical noise floor for astrometric planet detection.

Magnetic activity in stars creates features in the photosphere whose surface brightness is higher or lower than average. These features include the brighter faculae and the darker spots, the latter comprised of both penumbrae and umbrae Solanki (2003). As the star rotates and these features move across the visible portion of the stellar photosphere, they cause displacements in the observed photocenter, creating an astrometric jitter in time. Such displacements were demonstrated for Sun-like stars with a simple model by Lanza et al. (2008). These displacements mimic astrometric planet signals, and they must be mitigated for optimal performance.

To estimate the amplitude of the variations from stellar magnetic activity, Makarov et al. (2010) used maps of the bolometric surface intensity of the solar disk derived from the Mount Wilson Observatory magnetograms and intensity ratio images of the Sun to calculate the photocenter displacements. They found a maximum displacement of 0.56 milli solar radii (m$R_\odot$). Later, Lagrange et al. (2011) extended solar data to stars 10 pc away to see how significant this variability is compared to orbital reflex motion due to an exoplanet. They found that for stars with similar distributions of magnetic features as the Sun, the astrometric jitter from magnetic activity had an rms of 0.07 $\mu$as. These amplitudes are much smaller than those expected due to the orbit of an Earth-mass planet at a 1 au semimajor axis (averaging 0.3 $\mu$as along the equatorial plane), so they concluded that instrumental precision

---

[3] https://exoplanetarchive.ipac.caltech.edu/docs/counts_detail.html

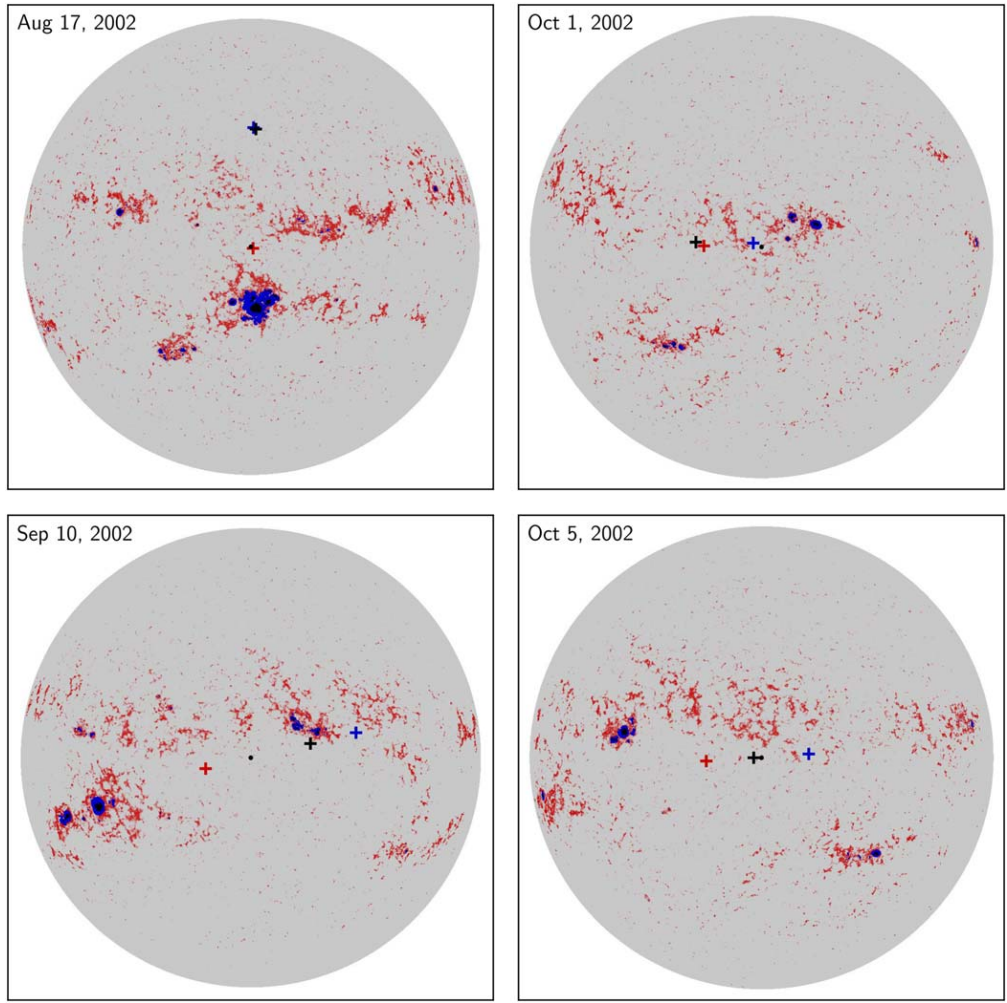

**Figure 1.** The shift in the photocenter on different days due to solar magnetic features. Distributions of bright faculae (red) and dark spot penumbrae (blue) and umbrae (black) are shown on the solar disk corresponding to the quiet Sun (gray), as used in the SATIRE-S model. The three crosses give the photocenter position according to shifts due to just faculae (red), just spots (blue), and both faculae and spots (black). Photocenter displacements from the center of the quiet Sun are scaled up by a factor of 1000 to be noticeable. The upper left panel shows a distribution dominated by dark spots repelling the photocenter, and the upper right panel shows a distribution dominated by bright faculae attracting the photocenter. The bottom panels show distributions with significant contributions from both faculae and spots, but in the bottom right panel these contributions happen to be nearly equal and opposite, canceling each other out and leaving almost no overall photocenter shift.

rather than magnetic activity is the limiting factor for planet detection in this case.

More recently, Morris et al. (2018) compared the effects of stellar magnetic activity with the precision of Gaia's astrometry. By simulating Gaia observations on stars, accounting for shifts in the photocenter due to just dark starspots, they found that Gaia's precision is likely sufficient to detect the astrometric effects of magnetic activity in nearby active stars.

In this paper, we investigate a multiwavelength strategy for stellar magnetic activity noise mitigation, to aid in astrometric detection of exoplanets. We use solar simulations from Shapiro et al. (2021) in five different passbands to leverage correlations in photocenter displacements across wavelengths. We neglect effects related to instrumental precision to demonstrate reducing the astrophysical noise floor due to just magnetic activity in Sun-like stars. In Section 2, we summarize the methodology for turning the solar simulations into astrometric measurements and compare the chosen passbands. In Section 3, we inject planet orbits into these simulations and test if we can recover the planet signal as a detection. In Section 4, we investigate multiwavelength noise mitigation and demonstrate its effectiveness in a sample case. Lastly, in Section 5, we show how different pairs of passbands reduce the lowest detectable planet mass with this approach to noise mitigation of stellar activity.

## 2. Solar Data

We first wish to model the astrometric noise due to stellar magnetic features in stars with levels of magnetic activity similar to the Sun. For these models, we use the results obtained by Shapiro et al. (2021), and we outline the most important steps in their analysis here for completeness.

These astrometric time series utilize the SATIRE-S model of the Sun Krivova et al. (2011), which derives the distribution of the magnetic features of the solar disk from daily solar magnetograms and intensity images. The model takes the solar equator to be at a $7\overset{\circ}{.}25$ angle with respect to the ecliptic. We project all motion of the photocenter along the $X$-axis, aligned with the solar equator, therefore treating the motion in 1D for simplicity. We choose this axis because it is the direction with the most day-to-day astrometric jitter; a full 2D analysis of the astrometric motion is outside the scope of this work. The photocenter is then calculated as a function of time as the intensity-weighted average of the $X$ coordinates of each pixel:

$$\overline{X}(t) = \frac{\sum_i F_i(t) X_i}{\sum_i F_i(t)}, \quad (1)$$

where the weight $F_i(t)$ is the irradiance of the $i$th pixel at time $t$. These pixel irradiances are calculated in terms of $F_i^k$ corresponding to the irradiance of the $i$th pixel due to just the $k$th type of magnetic feature. The SATIRE-S model considers bright faculae ($k=1$) and dark sunspot umbrae ($k=2$) and penumbrae ($k=3$), with $k=0$ representing the quiet Sun (i.e., no magnetic features). Therefore, the weights can be expressed in terms of the model's parameters as

$$F_i(t) = F_i^0(t) + \sum_{k>0} \alpha_i^k(t)(F_i^k(t) - F_i^0(t)), \quad (2)$$

where $\alpha_i^k(t)$ is the fractional coverage of the $i$th pixel by the $k$th type of magnetic feature, as calculated from the magnetograms in Yeo et al. (2014). The coordinate system is defined such that the origin is the photocenter of the quiet Sun, which requires that $\sum_i F_i^0(t) X_i = 0$.

In Figure 1, we show sample solar disks from the SATIRE-S model, with the resulting photocenters marked, to demonstrate how the photocenter shifts due to magnetic features according to Equation (1).

The $F_i^k(t)$ values in Equation (2) are calculated within a given passband, to give the photocenter within that passband, as

$$F_i^k(t) = \Delta\Omega \int_\lambda I_k^\epsilon(\lambda, \mu_i, t) \phi(\lambda) d\lambda \quad (3)$$

for energy intensities $I_k^\epsilon$ of the $k$th feature. For detectors that count photon intensities $I^\gamma$ instead, such as Gaia's CCD detectors, we can convert to energy intensities by multiplying by the photon energies to get $I^\epsilon(\lambda, \mu_i, t) = \frac{hc}{\lambda} I^\gamma(\lambda, \mu_i, t)$, where $\lambda$ is the central wavelength, with $h$ as Planck's constant and $c$ as the speed of light.

Here $\Delta\Omega$ is "the solid angle of the region on the solar surface that corresponds to one pixel when the magnetograms are obtained at 1 au from the Sun" (Shapiro et al. 2021). $I^\gamma(\lambda, \mu_i, t)$ is the intensity measured by the detector at wavelength $\lambda$ and given $\mu_i$, the cosine of the heliocentric angle at the $i$th pixel (i.e., the angle between the vector from that pixel to the observer and the vector normal to the solar surface at that pixel). $\phi(\lambda)$ is the transmission curve of the given passband.

The time series from Shapiro et al. (2021) include simulations for five different passbands. The first three of these correspond to the $G$, $B$, and $R$ passbands of the Gaia astrometry mission according to the Gaia DR2 revised passbands.[4] The fourth is the near-infrared passband of the Small-JASMINE mission planned for 2024 (Utsunomiya et al. 2014), which is treated as a top hat transmission profile between 1100 and 1700 nm. The fifth simulation corresponds to the Total Spectral Irradiance (TSI), which is the spectrally integrated energy flux (Kopp 2016).

Passbands were chosen to illustrate the wavelength dependence of the astrometric signal due to magnetic activity. The spectral transmission profiles for the Gaia passbands and the near-infrared Small-JASMINE passband are shown in Figure 2, along with the spectra used in the SATIRE-S model for the quiet Sun, the bright faculae, and the dark sunspot penumbrae and umbrae. These spectra are computed according to the work done by Unruh et al. (1999), and they are then used to produce the astrometric time series.

---
[4] https://www.cosmos.esa.int/web/gaia/iow_20180316

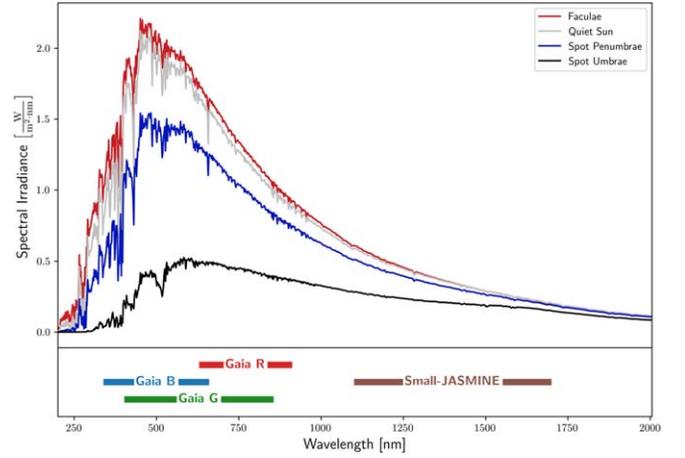

**Figure 2.** The spectral transmission profiles of the DR2 revised passbands for Gaia $G$, $B$, and $R$, and the expected passband for the upcoming Small-JASMINE mission. The range between their half maxima are shown as thick colored lines (bottom panel). Also plotted are the spectra used by the SATIRE-S model for the bright faculae, quiet Sun, and dark spot penumbrae and umbrae (upper panel). It can be seen by the overlap of these profiles with the different spectral bands that the amplitude of the astrometric jitter due to each feature will vary across bands, since relative contrast between features varies by band. Due to the close spectral proximity of the visible-light Gaia bands, we expect their astrometry to be more highly correlated with each other than with the near-infrared Small-JASMINE band. Additionally, since the relative contrast between magnetic features is lower at redder wavelengths, we expect motion due to stellar activity to be greater at bluer wavelengths.

The simulations of the astrometric signal in the five different passbands are taken with daily cadence from 1999 February 2 to 2014 August 1, corresponding to a 15.5 yr duration, spanning 1.4 solar activity cycles. The time series for the Gaia $B$ passband is shown in the left panel of Figure 3. The astrometric jitter is of order 0.5 m$R_\odot$, which is approximately the same amplitude as the Earth's gravitational effect on the Sun's center of mass.

We wish to inject reflex motion due to orbiting planets into these simulations and check whether the planet signal can be detected. These detections are difficult in the time domain but straightforward in the frequency domain. In the frequency domain, for a simulation duration much longer than the orbital period, the signal due to a circular orbit approaches a delta function at the orbit's frequency. Therefore, in such cases the planet can simply be detected as a spike in the amplitude spectrum of the astrometry simulation. Planets with nonzero eccentricity would contribute spikes at higher-frequency harmonics to this spectrum, so for simplicity we confine our investigation to circular orbits.

To convert from the time series to the amplitude spectrum, we use the discrete Fourier transform (DFT). Before applying this transformation, we first multiply the time series by the Hanning window, which goes to zero at the edges to attenuate the signal. Because the DFT assumes perfect periodicity, this windowing reduces spectral leakage at high temporal frequencies. The Hanning window is shown in Figure 3 according to its definition,

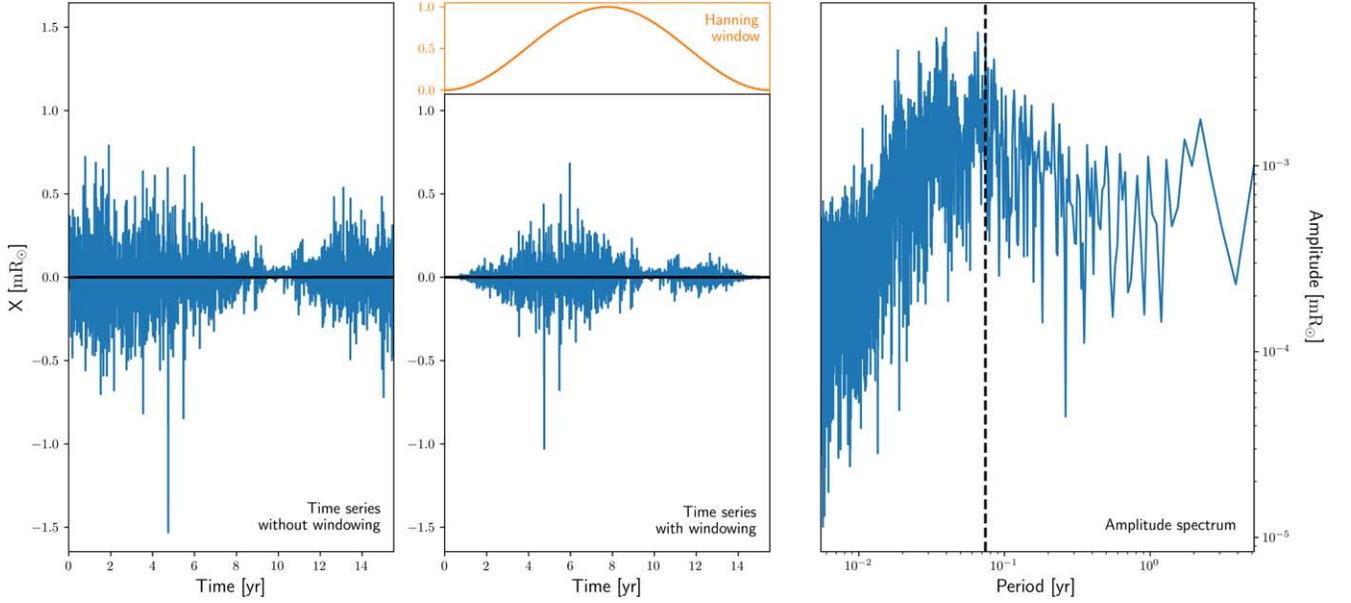

**Figure 3.** The Gaia *B* simulation of astrometric jitter in the solar photocenter due to faculae and spots. The raw time series of *X* displacement from the quiet Sun centroid is shown in the left panel. The Hanning window and its application to the time series is shown in the middle panel. Both time series include a solid line at $X = 0$. The right panel shows the amplitude spectrum derived from applying the discrete Fourier transform to the windowed time series. The vertical dashed line in the amplitude spectrum shows the solar rotation period of 27 days, where a bump in the spectrum is approximately centered.

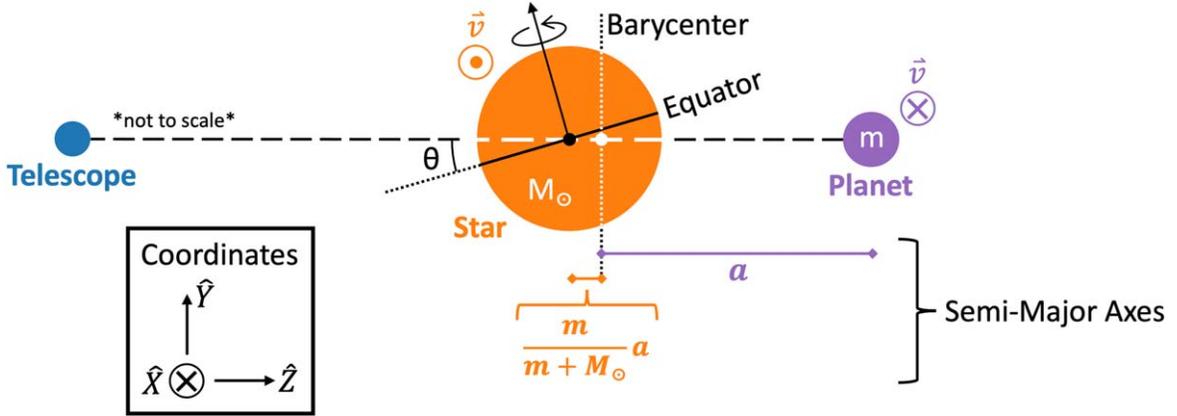

**Figure 4.** A diagram of the setup for the star-planet orbital system when $t = 0$ at transit. The telescope is shown on the left, with the host star and its orbiting planet along the line of sight. The star's equator is at the solar inclination $\theta = 7°\!.25$, and its axis of rotation is shown with an arrow normal to the equatorial plane. The planet's and star's tangential velocities $\vec{v}$ are shown perpendicular to the page, in the $\pm \hat{X}$ direction. The distance from the barycenter to the planet is the semimajor axis $a$, and from it we express the star's semimajor axis, $\frac{m}{m + M_\odot}a$. This result is then used to find the movement of the star's center of mass (and thus photocenter) due to reflex motion from the orbiting planet, as in Equation (7).

$$w(n) = 0.5 - 0.5 \cos\left(\frac{2\pi n}{N_d - 1}\right) \quad 0 \leqslant n \leqslant N_d - 1, \quad (4)$$

as a function of the *n*th data point for $N_d$ total data points. We then take the DFT of the windowed time series, defined as

$$A_\ell = \sum_{n=0}^{N_d - 1} a_n e^{-2\pi i n \ell / N_d} \quad 0 \leqslant \ell \leqslant N_d - 1, \quad (5)$$

where in this case $a_n$ corresponds to the *X* displacement of the *n*th data point, for $N_d$ total data points. An example of this process is shown in Figure 3, starting from the time series of the simulation in the Gaia *B* passband.

### 3. Planet Detection

We next outline our scheme for estimating the lowest detectable planet mass at a given semimajor axis with a Sun-like host star. To do so, we inject reflex motion due to a planet in circular orbit into the simulations of stellar astrometry at the five passbands. For simplicity we assume circular planetary orbits and project all reflex motion along the *X*-axis—equivalent to an edge-on orbit—so that we are only considering astrometry in one dimension. This assumption allows us to work in a near-worst-case scenario where the orbital motion is projected along the same axis as most of the astrometric jitter due to stellar activity. A schematic diagram of the setup is provided in Figure 4.

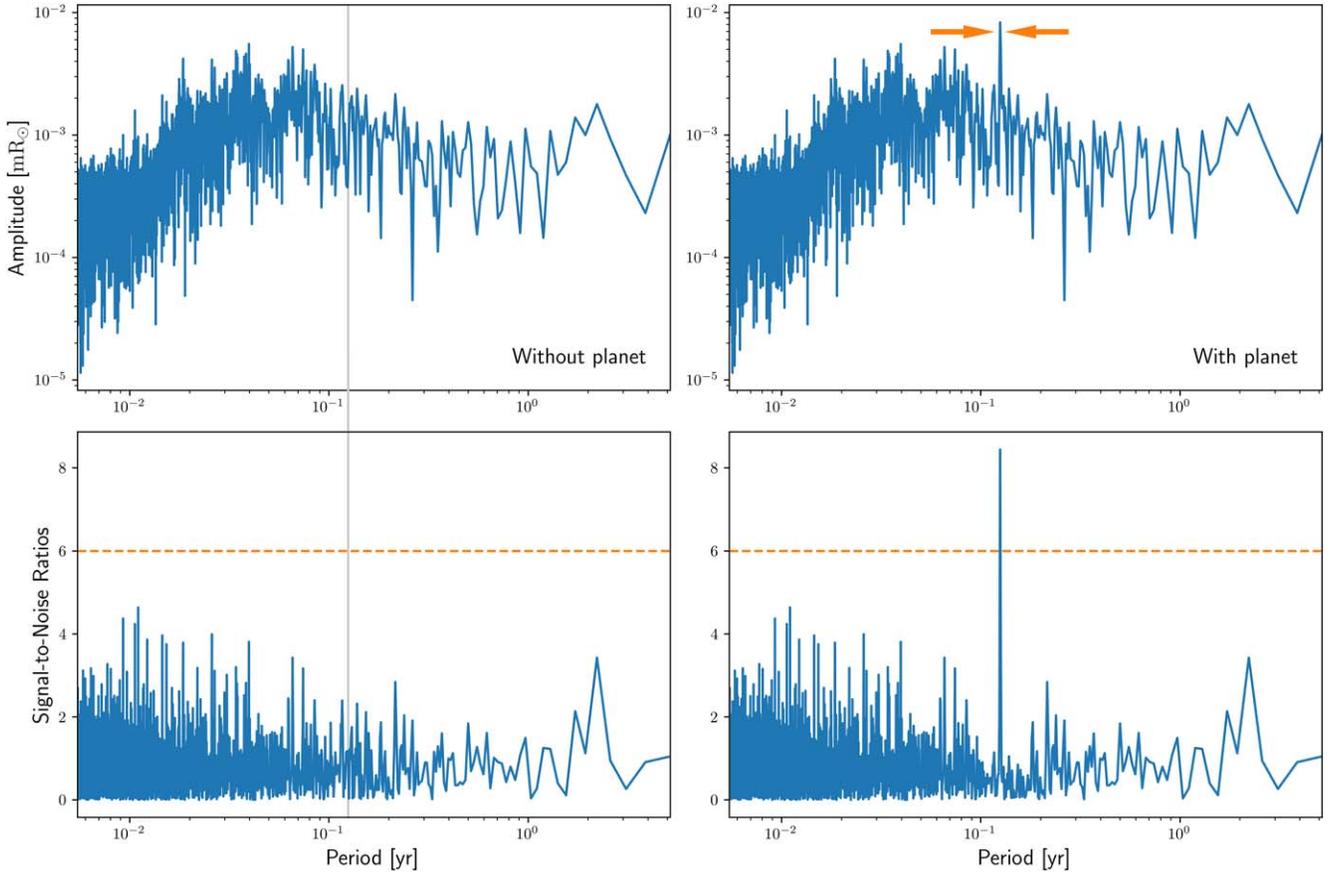

**Figure 5.** A comparison between a simulation in the Gaia *B* passband with and without a planet signal. The upper panels show the amplitude spectra of the simulations with just the stellar activity noise (upper left) and with an injected planet orbit for a 0.1 Earth-mass planet with a semimajor axis of 0.25 au (upper right). These amplitude spectra are derived from the methodology described in Section 2, with the injected planet orbit showing up in the upper right panel as a spike at the orbital period of $T = 0.125$ yr, obtained from Equation (6). The bottom panels show the signal-to-noise ratios derived from these amplitude spectra according to Equation (9). The $6\sigma$ detection threshold is shown as a dashed horizontal line. For the case without the planet, there are no signals detected (bottom left). For the case with the planet (bottom right), the signal-to-noise only surpasses the threshold at the orbital period of the planet, which is then flagged as a detection (with the spike indicated by arrows in the upper right panel). This orbital period is also marked by a vertical line in the left panels, but there is no planet signal there to detect.

For a given semimajor axis $a$, the planet's orbital period can be calculated from Kepler's Third Law as

$$T = 2\pi\sqrt{\frac{a^3}{GM_\odot}}, \quad (6)$$

with solar mass $M_\odot$ and gravitational constant $G$ (Carroll & Ostlie 2007). We choose $1\,M_\odot$ for the host star's mass because our simulations use solar data and therefore apply to the levels of magnetic activity found in Sun-like stars. We can simulate this one-dimensional reflex motion as a function of the time $t$ with the following equation for photocenter displacement:

$$\Delta X(t) = \left[\frac{m}{m + M_\odot}a\right]\sin\left(\frac{2\pi}{T}t\right), \quad (7)$$

where $m$ is the planet's mass. The amplitude of the reflex motion is the semimajor axis of the host star's orbit, $\frac{m}{m+M_\odot}a$. Since the star is in circular orbit about the center of mass of the star-planet system, our formula for displacement due to the planet's orbit is this semimajor axis modulated by a sinusoidal temporal function with period $T$. Without loss of generality we restrict the phase to be 0, meaning $t = 0$ physically corresponds to when the planet is transiting (i.e., closest to or farthest from us).

To convert from Equation (7) to angular units, we can divide by the distance to the system in the small angle approximation. However, since we are neglecting instrumental precision for this analysis, the fundamental limit we wish to examine is given by actual oscillations in physical space. Therefore, we just use Equation (7) for the physical displacements.

The motion given by Equation (7) is the same at every wavelength, since it corresponds to physical motion of the host star's center of mass. We can now generate temporal amplitude spectra of a system with both the noise source (stellar activity) and the signal (orbital motion). To do so, the orbital motion given by Equation (7) is added to the time series of photocenter motion due to stellar activity. We then apply the DFT methodology described in Section 2, multiplying both the signal due to stellar activity and orbital motion by the Hanning window and then taking the transform. A sample amplitude spectrum in the Gaia *B* passband that includes an injected planet orbit is shown in Figure 5.

Our goal is to determine, for a given semimajor axis (and thus orbital period, by Equation (6)), the lowest planet mass that can be detected with the simulated astrometric jitter due to stellar activity. We treat this problem in the limit of arbitrarily small angular resolution to study the noise floor due to just magnetic activity. To determine this detection curve, we implement a simple detection algorithm that searches for the planet signal as a spike in the temporal amplitude spectrum above a $6\sigma$ threshold, i.e., a signal-to-noise ratio of 6.

For converting from the amplitude spectrum to the signal-to-noise ratios with respect to period, we first de-trend the amplitude spectrum by subtracting the median at each point. This median is computed over a window of radius $w = 30$ points on each side of the data point in question, excluding the data point itself. This means that for a set of $N_d$ points $\{x_p\}$ for $0 \leqslant p \leqslant N_d - 1$, we take the detrended data points $\{\hat{x}_p\}$ to be

$$\hat{x}_p = x_p - \text{Median}[x_{p-w}, \ldots, x_{p-1}, x_{p+1}, \ldots, x_{p+w}], \quad (8)$$

where we truncate this window for points that are $w_0 < w$ points away from the left (right) edge of the list to include only $w_0$ points on the left (right) side. We then take the signal $S$ to be the detrended value at the point ($S = \hat{x}_p$), and we take the noise to be the standard deviation over the same window described above. This gives signal-to-noise ratios (S/Ns) of

$$\text{S/N} = \frac{\hat{x}_p}{\text{STD}[\hat{x}_{p-w}, \ldots, \hat{x}_{p-1}, \hat{x}_{p+1}, \ldots, \hat{x}_{p+w}]}. \quad (9)$$

Because the resolution of data points is low at high periods, we count signals that are within 3 points of each other as the same detection, since these are likely delta functions with some finite width due to the low resolution. This complete detection scheme is shown for the Gaia $B$ simulation in Figure 5, where the injected planet orbit shows up in the signal-to-noise plot as a detection with S/N > 6.

Noise due to stellar activity varies across periods, so this detection scheme requires higher amplitude signals at some periods than others. For example, in Figure 5, the noise is largest around 0.1 yr, and therefore a larger amplitude signal is required for a detection, despite the constant $6\sigma$ threshold.

As seen in Figure 5, there are many outliers at lower periods, likely because the stellar activity is in reality non-Gaussian. We are nevertheless treating it as Gaussian noise for simplicity and to allow us to compare algorithms. Note that these outliers do not result in any false positives for our data set, because none of them pass the $6\sigma$ detection threshold. Evaluating real detection thresholds would require a more careful treatment of the statistics of the data. For elliptical orbits and irregular sampling, techniques such as those used in radial-velocity detection (e.g., the Lomb-Scargle Periodogram; VanderPlas 2018) would be employed, and more sophisticated detection algorithms used. Such approaches would likely only change the detectable planet threshold by a factor of a few.

## 4. Noise Mitigation

Using the detection scheme described in Section 3, for a planet with semimajor axis of 1 au, we find that the lowest detectable planet mass is 0.01 $M_E$ in the Small-JASMINE passband. In contrast, in the Gaia $G$ band, we can detect planets at 1 au down to masses of 0.04 $M_E$. These numbers are obtained by successively increasing the planet mass injected into the simulation and noting the lowest mass where the planet signal is first detectable. (See Section 5 for details.)

These planet masses are already small in the single-band case because we are treating this problem in the limit of infinite telescope resolution and rapid (daily) cadence; planet masses this low would not be detectable by Gaia astrometry, for example. Therefore, it is still in our interest to find techniques for noise mitigation to reduce the noise floor due to stellar magnetic activity. Motivated by how stellar magnetic features contribute to photometry at different amounts in different passbands, as demonstrated in Figure 2, we explore how leveraging cross-wavelength correlations can help reduce this noise floor.

We begin by comparing the correlations in astrometric jitter due to magnetic activity across each pair of passbands. At a given time, the location of these magnetic features on the solar disk is the same at all wavelengths, but the amount that each feature biases the photocenter will vary by wavelength because the contrasts of each feature are wavelength dependent. Therefore, we expect some scaling factor between the $X$ displacement in one passband and the $X$ displacement in another. These correlations and scaling factors are shown in Figure 6 for our data set, where at each epoch in the simulation, we have plotted the displacement in one passband against the displacement in the other.

We use the Pearson correlation coefficient $r$ to measure correlation between two data samples $x$ and $y$, with mean values $\bar{x}$ and $\bar{y}$, respectively, and each with $N_d$ data points. This value is determined by the following equation:

$$r = \frac{\sum_{i=1}^{N_d}(x_i - \bar{x})(y_i - \bar{y})}{\sqrt{\sum_{i=1}^{N_d}(x_i - \bar{x})^2}\sqrt{\sum_{i=1}^{N_d}(y_i - \bar{y})^2}}. \quad (10)$$

Some pairs of passbands are well correlated (i.e., $r$ is close to 1), whereas others, such as pairings between Gaia bands and the Small-JASMINE (SJ) passband, are poorly correlated (indicated by the cloudiness of the points). These patterns are partly explained by the spectral proximity of the passbands because the distributions and relative contrasts between magnetic features are most similar at nearby wavelengths. The trends in correlation can also be attributed to the types of magnetic features that dominate the photocenter motion at different wavelengths. In the Gaia passbands, bright faculae play a dominant role for periods longer than a solar rotation, since few starspots live that long but most faculae do. In the near-infrared region, on the other hand, facular contrast is far lower, so photocenter motion is dominated by starspots even at longer timescales, despite few of them lasting sufficiently long. Comparing across these two regimes, which are dominated by different types of magnetic features, therefore leads to low correlations in apparent motion due to stellar activity. These traits, and in particular the excellent correlation between the brightness of faculae observed at visible wavelengths, is demonstrated in Unruh et al. (1999).

Similarly, only some pairs of passbands have lines of best fit with slope $\mathcal{M}$ far from 1. These are pairs where the relative contrast between magnetic features is very different between the two passbands. For example, the pair of Small-JASMINE and Gaia $B$ gives rise to the farthest slope from 1, at $\mathcal{M} = 0.31$. Referring to Figure 2, we see that Gaia $B$ has high contrast between different types of features compared to the near-infrared range, where every type of magnetic feature approaches similar levels of irradiance. Photocenter motion due to stellar magnetic activity is therefore much greater in Gaia $B$ than in Small-JASMINE, accounting for the slope far from 1.

Generally, we see that some pairs of passbands have motion due to stellar activity that is very well correlated with nonunity slope, with less motion in the photocenter for the redder band. For one such pair, labeled passbands 1 and 2, let $A_1$ and $A_2$ be their amplitude spectra (the set of amplitudes at each period), respectively, with scaling factor $\mathcal{M}_{2,1} \sim A_2/A_1$. This $\mathcal{M}$ is really the slope of the line of best fit between the *X* displacements in passbands 1 and 2.

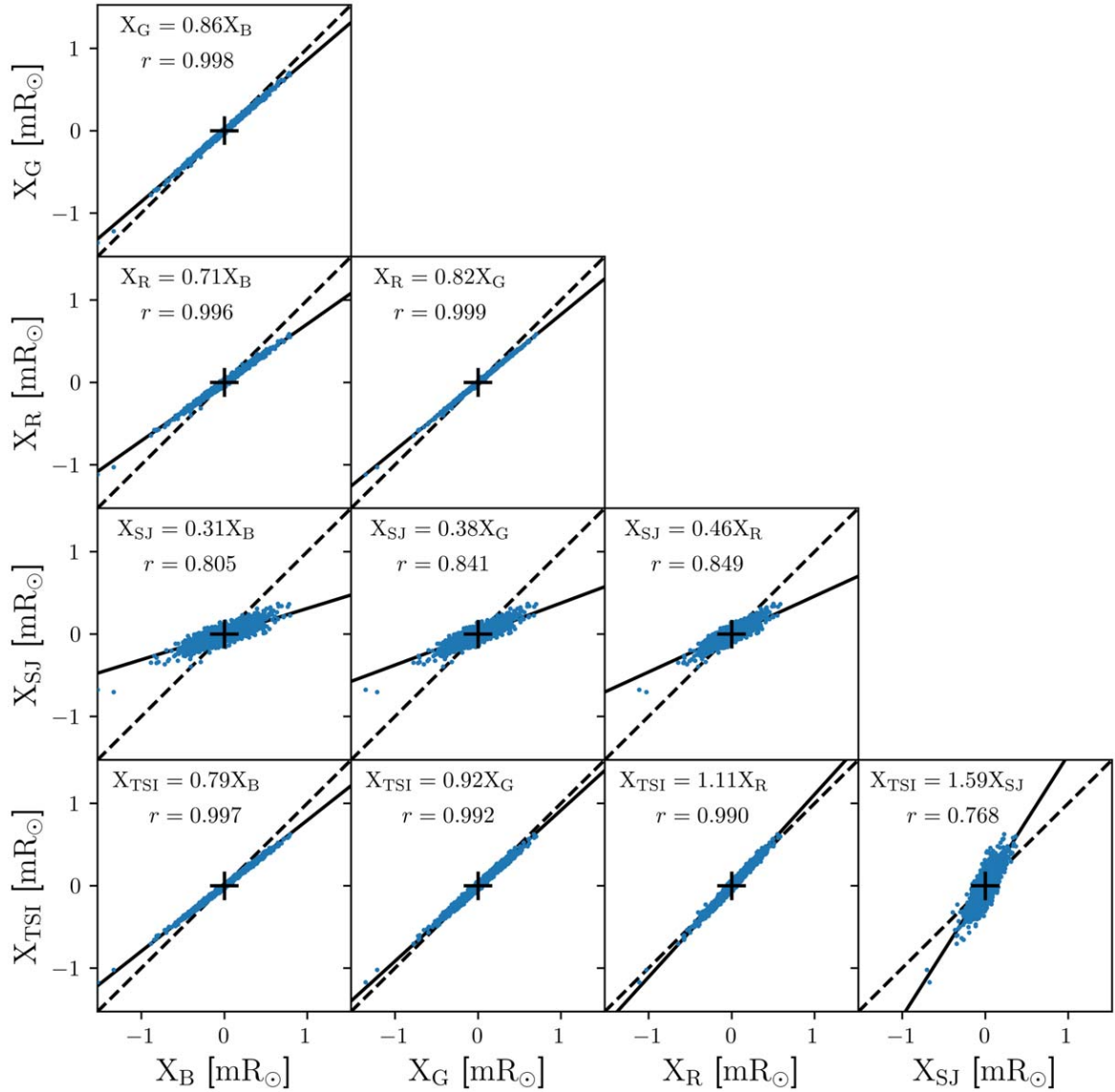

**Figure 6.** The correlation between passband pairs in *X* displacement due to stellar magnetic activity. Each data point corresponds to a given epoch in the solar simulation. The dashed lines have a slope of 1, corresponding to equal displacement in both passbands. The solid line shows the line of best fit for the data points of that pair of passbands, where the slope, $\mathcal{M}$, is the scaling factor between the amplitudes of one passband and the other, annotated at the top of each plot. Beneath that is the correlation *r* of the data points as defined in Equation (10).

Next we inject a common planet signal into both simulations, which should have the same amplitude for both passbands, since the physical motion of the host star's center of mass is independent of wavelength. We denote these new amplitude spectra with both the stellar activity-induced motion and the orbital motion as $\widehat{A}_1$ and $\widehat{A}_2$.

By scaling the amplitudes in one passband to match the amplitude of the motion due to stellar activity in the other, the planet signal will now have a different amplitude in each passband. Subtracting one spectrum from the other, we attenuate most of the stellar activity noise but keep the residual planet signal from rescaling. Accordingly, we take our new amplitude spectrum to be

$$\widehat{A} = \widehat{A}_2 - \mathcal{M}_{2,1}\widehat{A}_1. \quad (11)$$

For appropriate passband pairs, $\widehat{A}$ should have a higher signal-to-noise ratio at the planet's orbital period compared to $\widehat{A}_1$ and $\widehat{A}_2$. This process could therefore allow for lower detectable planet masses, using information from two passbands instead of just one. An example of this noise mitigation process is shown in Figure 7 for the Gaia *R* and Gaia *B* passbands. We see that the planet signal is not detected in each individual passband, but combining them with this scaling and subtraction method allows for a planet detection.

## 5. Results

To study the effectiveness of this noise mitigation strategy, we plot curves of the lowest detectable planet mass versus the semimajor axis of the planet's orbit, shown on the diagonal panels of Figure 8 for each passband. To estimate the lowest detectable planet mass at a given semimajor axis, we begin with a low planet mass of $1 \times 10^{-3}\,M_E$ and check if there is a planet detection using the methodology described in Section 3. If there is, we return this planet mass as the lowest detectable one. If not, we increase the planet mass slightly and repeat this process, until we find a detection.

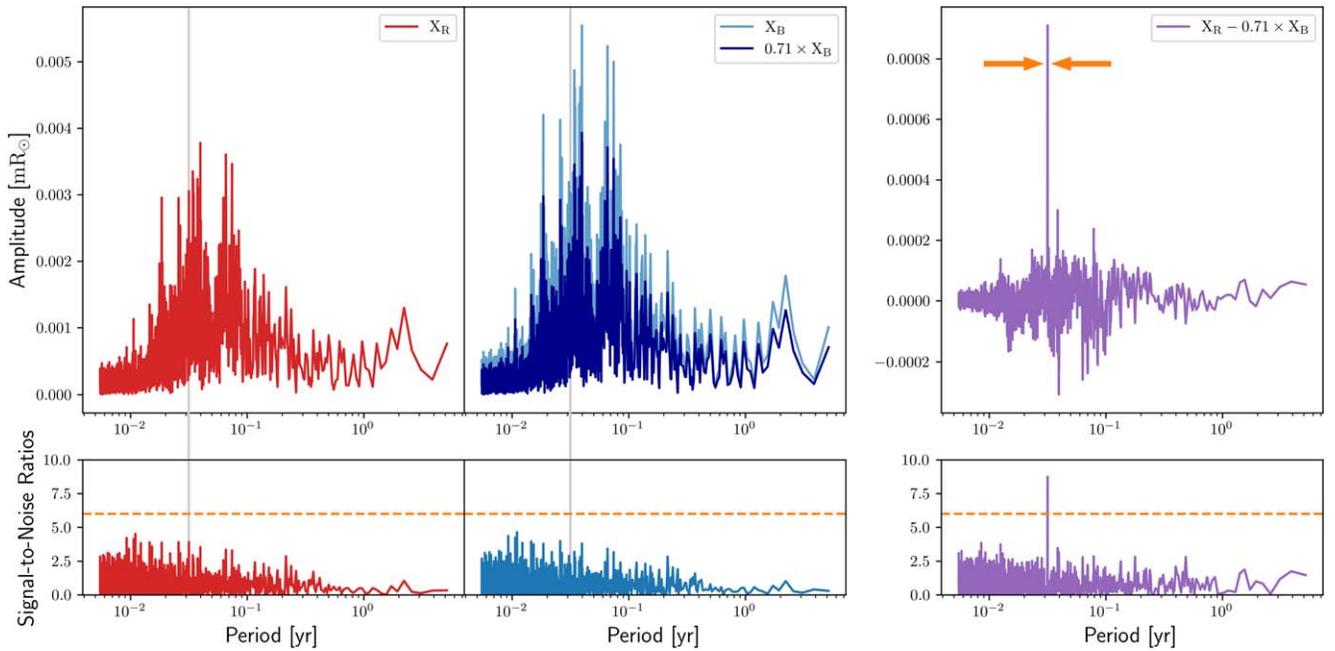

**Figure 7.** Multiwavelength noise mitigation using the Gaia *R* and Gaia *B* passbands for a case where a planet cannot be detected in any single passband. A planet of mass $0.1\,M_E$ is injected into the simulations at semimajor axis 0.1 au. The amplitude spectrum and corresponding signal-to-noise plot is shown for the Gaia *R* simulation (left panels), with a solid vertical line at the planet's orbital period. The same respective data is plotted for Gaia *B* (middle panels), but also with the amplitude spectrum multiplied by the scale factor, to scale the *X* displacements down to the same amplitude as in the Gaia *R* passband. (This factor can be found in the second row, first column of Figure 6.) In both single-passband plots, the planet is not actually detected—the S/N does not cross the $6\sigma$ threshold at the planet's orbital period. However, subtracting the scaled down $X_B$ amplitudes from the $X_R$ amplitudes significantly reduces the noise due to stellar activity and amplifies the planet signal relative to it (right panels). Only after this subtraction is the planet signal flagged as a detection, with the spike indicated by the arrows in the upper right panel.

The resulting detection curves trend downwards as the semimajor axis increases, since the amplitude of orbital reflex motion in the host star scales linearly with the planet's semimajor axis. We see that even without noise mitigation, some passbands are better than others for planet detection. At 1 au, we find that across all passbands, the median detectable planet mass for a Sun-like star is $0.039\,M_E$.

Note that Small-JASMINE shows the lowest detectable planet masses for single-band astrometry. Referring to Figure 2, we see that the magnetic features and the quiet Sun have similar spectral profiles in the near-infrared range of Small-JASMINE, compared to the other passbands. This means the astrometric noise due to magnetic activity will be much smaller in this passband, allowing for better detections. However, when taking into account observational limitations, stars will be dimmer and worse-resolved in the near-infrared range, which may balance out the benefits from this quieter passband. Similar results have been expected for Doppler spectroscopy, where near-infrared observations may be less sensitive to stellar activity (Cale et al. 2018).

We next compare these unmitigated detection curves with detection curves after the noise mitigation strategy described in Section 4. A drop in the detection curve after noise mitigation would indicate that our approach helps with planet detection. In Figure 8, each panel compares a single-band detection curve with the results of noise mitigation between one passband and another. Some pairs of passbands, such as Gaia *G* and TSI, do not help with detectable planet mass. The pair TSI and Small-JASMINE actually harms planet detection significantly, likely because the correlation in stellar activity-induced jitter is the lowest for this pair of bands ($r = 0.768$). Most curves, however, show significant improvement after multiwavelength noise mitigation.

To quantify the effectiveness of noise mitigation, we report a score for each pair of passbands. This score is found by taking the median of the unmitigated curve divided by the noise-mitigated curve, in the range of $0.7 \to 1.5$ au, which approximately matches the habitable zone of planets. For a single passband labeled 1, we denote the lowest detectable planet mass values in that passband as $\{m_a^1\}$, where $a$ indexes the semimajor axis values. For the corresponding lowest detectable planet masses $\{m_a^{1,2}\}$ after noise mitigation using a second passband, labeled 2, we compute this pair's score as

$$\text{Score}_{1,2} = \text{Median}\left[\frac{m_l^1}{m_l^{1,2}}, \frac{m_{l+1}^1}{m_{l+1}^{1,2}}, \ldots, \frac{m_{u-1}^1}{m_{u-1}^{1,2}}, \frac{m_u^1}{m_u^{1,2}}\right], \quad (12)$$

where $l$ and $u$ are the indices of the semimajor axis values closest to 0.7 and 1.5 au, respectively. According to our definition for these scores, we see that for planet detection,

$$\begin{cases} \text{Score} > 1 & \to \text{helps.} \\ \text{Score} = 1 & \to \text{does not help or hurt.} \\ \text{Score} < 1 & \to \text{hurts.} \end{cases} \quad (13)$$

These scores are reported in the bottom left of each panel in Figure 8. The exact values of the scores depends on the details of the activity during the modeled solar cycles, but in considering the entire habitable zone, we should average out most of the effects from these details. We find that the best pair of passbands for this kind of noise mitigation is Gaia $G$ and Gaia $R$, with a score of 5.84. At 1.0 au, mitigation with these passbands corresponds to a detection limit of $4.7 \times 10^{-3} M_E$, reducing this detectable mass limit by almost a factor of 10 relative to the single-band case. The best improvement seen in lowest detectable planet mass across all semimajor axes is a factor of 19 for this same pair, at a planet semimajor axis of 1.8 au.

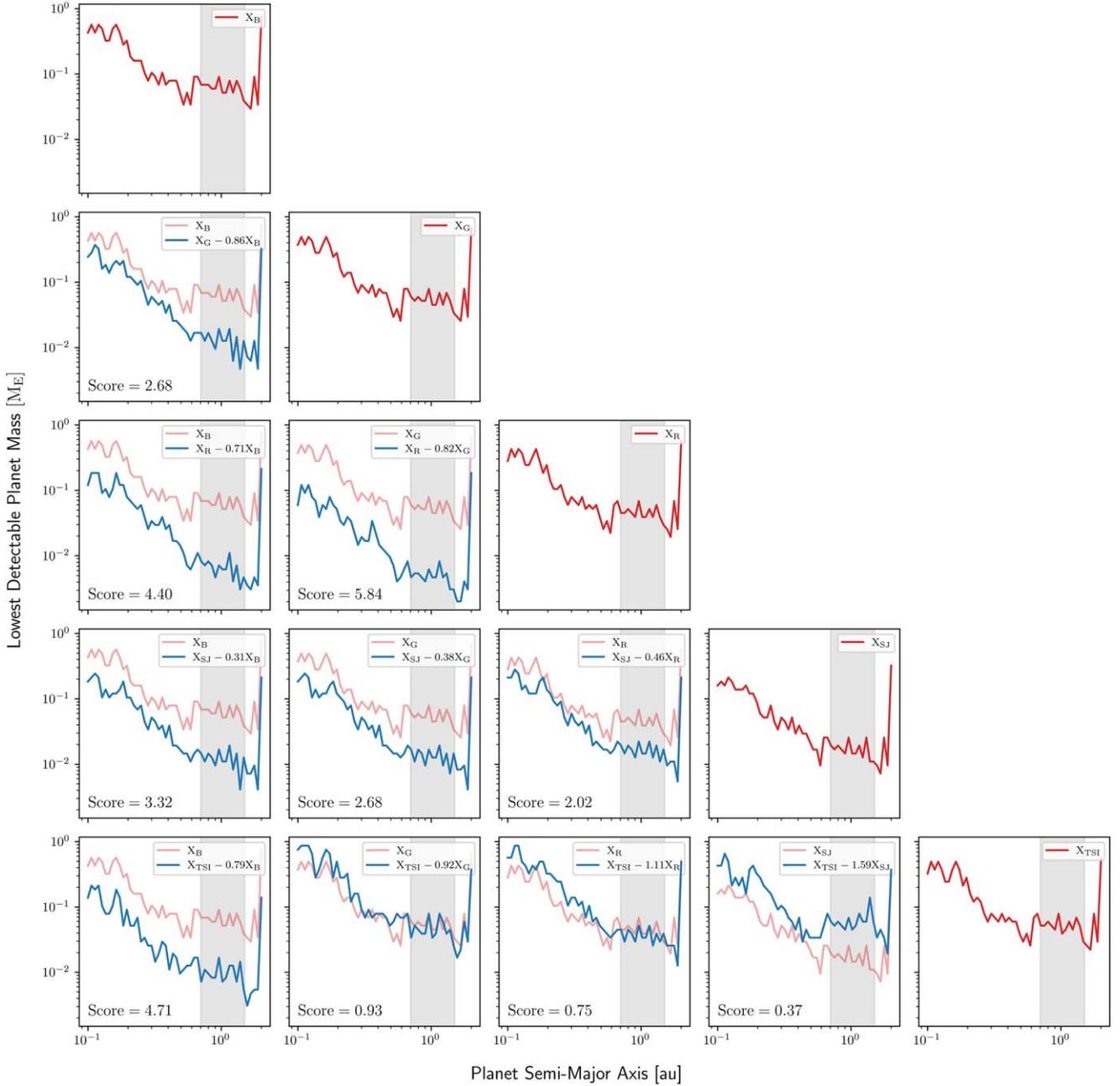

**Figure 8.** The lowest detectable planet mass curves, comparing between the single-band and multiband methods. We plot these curves for semimajor axes between 0.1 and 2.0 au, with 50 points spaced equally on a log scale. For scanning across planet masses until we find a detection, we also search on a log scale. Diagonal plots show results for single-band astrometry, and these curves are shown in each panel of that passband's column for comparison. The legend in the top right of each panel shows first the passband used as the single-band treatment, and second the subtraction done for noise mitigation according to Equation (11), with the scaling values taken from Figure 6. The sophisticated small-scale features of the curves are highly dependent on the chosen interval of solar time, since they are determined by the locations of the peaks in the solar amplitude spectrum. The general trends we are interested in should nevertheless be robust, regardless of the time series' specific interval. Noise mitigation scores for each pair of passbands is reported in the bottom left of the panel corresponding to that pair, calculated according to Equation (12). Note that these scores are computed over a relatively narrow range of planet semimajor axes, highlighted in each panel, between 0.7 and 1.5 au.

To better understand the relationship between the chosen pair of passbands and the noise mitigation score, we look at the slopes $\mathcal{M}$ (i.e., the scaling factors) and the correlations $r$ reported in Figure 6 between each pair of passbands. In Figure 9, we plot the scores for each pair of passbands as a function of the distance from 1 of both the slope and the correlation between the two passbands. We interpolate between points to plot contours between the sparse data points. We find that generally, pairs of passbands that are highly correlated and with slope far from 1 are the best choice for noise mitigation, as expected. The high correlation means that scaling up the motion in one passband to match the other and then subtracting should successfully attenuate stellar activity-induced noise at all periods. The slope far from 1 means that when we scale up the planet signal that originally had equal amplitude in both spectra, subtracting one spectrum from the other leaves a significant residual planet signal to detect.

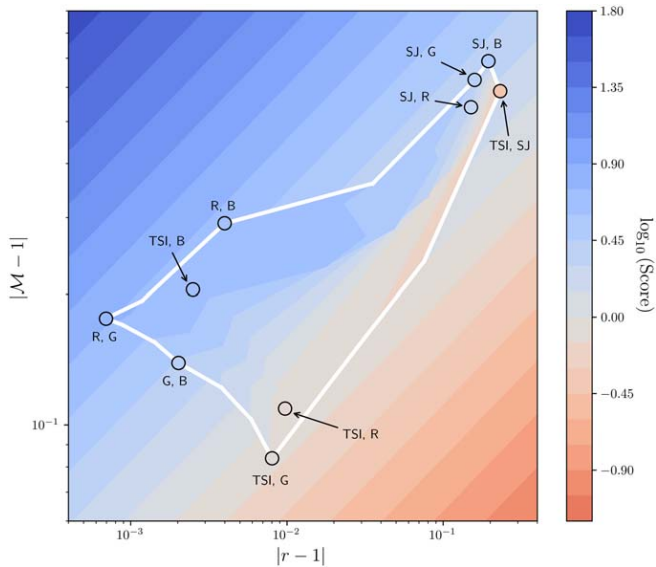

**Figure 9.** The noise mitigation scores of each pair of passbands, according to Equation (12). We plot $\log_{10}$(Score) vs. the slope's distance from 1 ($|\mathcal{M} - 1|$) and the correlation's distance from 1 ($|r - 1|$). Because we take the log of the slope, colors corresponding to positive values show improvement (blue), and colors corresponding to negative values show that planet detection became more difficult (red). The 10 data points are plotted as circles filled in with the color of their score and outlined in black, labeled by the passband names for the corresponding pair. The contours plotted within the boundaries of the white border are derived from linear interpolation of a nonuniform grid for the 10 scores, implemented with NumPy's tricontourf function. Outside these bounds, we plot the contours for our functional form from Equation (14) to show how the scores approximately extrapolate, although it is less reliable than the interpolation inside the white border.

We also fit power laws to the slope and correlation in order to extrapolate the contours past our data points. We take the score to be of the functional form

$$\text{Score} = C \, |r - 1|^\alpha \, |m - 1|^\beta, \quad (14)$$

for fit parameters $\alpha$, $\beta$, and C. Fitting with SciPy's implementation of nonlinear least squares, the resulting parameter values and their errors are as follows.

The result of this fit is plotted in the background of Figure 9. Because of the clear trend in favor of high correlation and slope far from 1 for higher scores, we conclude that, for this noise mitigation technique, passband pairs should be chosen such that the astrometric jitter (due to magnetic activity) is well correlated in both bands; but with a scaling factor that is far from 1 between the displacements in each passband.

### 6. Conclusions

We have demonstrated that for Sun-like stars, using multi-wavelength noise mitigation can successfully attenuate noise due to stellar magnetic activity, relative to the reflex motion induced in that star by an orbiting planet. For planet detection, this noise mitigation therefore reduces the noise floor due to stellar activity. Around the habitable zone (defined above as $0.7 \rightarrow 1.5$ au), with an appropriate pair of passbands, this corresponds to improving planet mass detection with stellar astrometry by a factor of up to 5.84.

**Table 1**
The Values Found for the Parameters in Equation (14) Using Least Squares Fitting

| Parameter | Value | Error |
|---|---|---|
| $\alpha$ | −0.47 | $8.12 \times 10^{-3}$ |
| $\beta$ | 1.27 | $1.27 \times 10^{-1}$ |
| C | 1.86 | $1.67 \times 10^{-1}$ |

We found that the pair of passbands for which this strategy is most effective are pairs with high correlation and a significant scaling factor between activity-induced apparent motion in the two passbands. Pairs of passbands where the relative contrasts between types of magnetic features are very different were the ones with scaling factor farthest from 1. This explains why, among the visible bands, pairing Gaia *R* with Gaia *B* or Gaia *G*, both of which have relative contrasts similar to each other but different from Gaia *R*, works better than pairing Gaia *B* with Gaia *G*.

We also see that correlation is lowest when pairing visible passbands with the near-infrared passband. This pattern can be explained by the spot-dominated nature of the near-infrared region compared to the faculae-dominated visible region. Since spots are much more short-lived than faculae, pairing passbands across these two regimes leads to low correlation. On the other hand, close spectral proximity between neighboring passbands means the relative contributions of dark spots and bright faculae remain largely unchanged between the two bands, giving rise to high correlations. Selecting pairs of passbands according to these observations, so that they are highly correlated with a scaling factor far from 1, would help with noise reduction of stellar magnetic activity. Our results indicate that two visible-light passbands, with different levels of relative contrasts between magnetic features, may be the best candidates for such missions.

In our analysis, we assumed circular orbits for the planets with entirely edge-on motion. For more general elliptical orbits, we expect spikes in the spectrum at higher-frequency harmonics, and the results may increase the lowest detectable planet masses. However, this assumption should only not significantly affect the relative scaling between single-band and multiband astrometry (i.e., the score of each pair of passbands).

Another relevant assumption was taking the orbital orientation to be such that the planet- and stellar-activity-induced motion was almost maximally aligned, meaning we were treating a near-worst-case scenario. Other features of the star itself, such as inclination and metallicity, may complicate planet detection and the mass curves. Their effects on solar data have been explored by Sowmya et al. (2021). These considerations again should not significantly impact the reduction factor from noise mitigation, despite their effects on individual detection curves.

The SATIRE-S model used for the astrometry simulations derived its distributions of stellar magnetic features from solar magnetograms and intensity ratio images, meaning our results apply to Sun-like stars with similar levels of magnetic activity. Stars of other stellar types have different magnetic activity signatures (Oláh et al. 2016), and in particular late-type and young stars may be more photospherically active.

Additionally, our study assumed a telescope with infinite resolution and no fundamental noise, so that we could focus on the noise floor due to just stellar magnetic activity. In reality, the lowest detectable planet masses we found will be much lower than planet masses found by actual astrometry missions, due to the observational limitations on astrometric precision. Considerations about spectral radiance would also need to factor into a choice of filters for a mission.

As searching for exoplanets via stellar astrometry becomes more viable, we recommend that missions with this objective consider performing astrometry in more than one passband, to allow for mitigation of noise due to stellar activity. Ideally, these passbands would be chosen to be well correlated (e.g., close spectral proximity) with a scaling factor for astrometric jitter far from 1. This way, noise induced by stellar magnetic activity should not be a significant barrier to planet detection with astrometry of Sun-like stars.

This research was largely inspired by discussions with Peter Graham, Surjeet Rajendran, and Michael Fedderke about sources of astrometric noise in the context of gravitational wave detection. Our work benefited as well from discussions with Monica Bobra and Meng Jin about Solar Dynamics Observatory data. The entire Gemini Planet Imager group provided an incredibly supportive and collaborative environment within which to conduct this research; thank you especially to Eric Nielsen and Lea Hirsch for their mentorship and advice at group meetings and beyond. We also wish to thank the Orbitize team, and particularly Sarah Blunt and Jason Wang, for their helpful contributions regarding orbit-fitting. Finally, we thank the Stanford Physics Department's Undergraduate Research Program for funding this research.

This project has received funding from the European Research Council (ERC) under the European Unions Horizon 2020 research and innovation program (grant agreements No. 695075 and 715947).

*Software:* IPython (Perez & Granger 2007), NumPy (Harris et al. 2020), SciPy (Virtanen et al. 2020), Matplotlib (Hunter 2007).

## ORCID iDs

Avi Kaplan-Lipkin https://orcid.org/0000-0002-8756-0347
Bruce Macintosh https://orcid.org/0000-0003-1212-7538
Krishnamurthy Sowmya https://orcid.org/0000-0002-3243-1230
Alexander Shapiro https://orcid.org/0000-0002-8842-5403
Natalie Krivova https://orcid.org/0000-0002-1377-3067
Sami K. Solanki https://orcid.org/0000-0002-3418-8449